\documentstyle[prd,aps,12pt,epsf,psfig]{revtex}
\begin{document}

\baselineskip=7mm
\def\ap#1#2#3{           {\it Ann. Phys. (NY) }{\bf #1} (19#2) #3}
\def\arnps#1#2#3{        {\it Ann. Rev. Nucl. Part. Sci. }{\bf #1} (19#2) #3}
\def\cnpp#1#2#3{        {\it Comm. Nucl. Part. Phys. }{\bf #1} (19#2) #3}
\def\apj#1#2#3{          {\it Astrophys. J. }{\bf #1} (19#2) #3}
\def\asr#1#2#3{          {\it Astrophys. Space Rev. }{\bf #1} (19#2) #3}
\def\ass#1#2#3{          {\it Astrophys. Space Sci. }{\bf #1} (19#2) #3}

\def\apjl#1#2#3{         {\it Astrophys. J. Lett. }{\bf #1} (19#2) #3}
\def\ass#1#2#3{          {\it Astrophys. Space Sci. }{\bf #1} (19#2) #3}
\def\jel#1#2#3{         {\it Journal Europhys. Lett. }{\bf #1} (19#2) #3}

\def\ib#1#2#3{           {\it ibid. }{\bf #1} (19#2) #3}
\def\nat#1#2#3{          {\it Nature }{\bf #1} (19#2) #3}
\def\nps#1#2#3{          {\it Nucl. Phys. B (Proc. Suppl.) } {\bf #1} (19#2) #3} 
\def\np#1#2#3{           {\it Nucl. Phys. }{\bf #1} (19#2) #3}

\def\pl#1#2#3{           {\it Phys. Lett. }{\bf #1} (19#2) #3}
\def\pr#1#2#3{           {\it Phys. Rev. }{\bf #1} (19#2) #3}
\def\prep#1#2#3{         {\it Phys. Rep. }{\bf #1} (19#2) #3}
\def\prl#1#2#3{          {\it Phys. Rev. Lett. }{\bf #1} (19#2) #3}
\def\pw#1#2#3{          {\it Particle World }{\bf #1} (19#2) #3}
\def\ptp#1#2#3{          {\it Prog. Theor. Phys. }{\bf #1} (19#2) #3}
\def\jppnp#1#2#3{         {\it J. Prog. Part. Nucl. Phys. }{\bf #1} (19#2) #3}

\def\rpp#1#2#3{         {\it Rep. on Prog. in Phys. }{\bf #1} (19#2) #3}
\def\ptps#1#2#3{         {\it Prog. Theor. Phys. Suppl. }{\bf #1} (19#2) #3}
\def\rmp#1#2#3{          {\it Rev. Mod. Phys. }{\bf #1} (19#2) #3}
\def\zp#1#2#3{           {\it Zeit. fur Physik }{\bf #1} (19#2) #3}
\def\fp#1#2#3{           {\it Fortschr. Phys. }{\bf #1} (19#2) #3}
\def\Zp#1#2#3{           {\it Z. Physik }{\bf #1} (19#2) #3}
\def\Sci#1#2#3{          {\it Science }{\bf #1} (19#2) #3}

\def\n.c.#1#2#3{         {\it Nuovo Cim. }{\bf #1} (19#2) #3}
\def\r.n.c.#1#2#3{       {\it Riv. del Nuovo Cim. }{\bf #1} (19#2) #3}
\def\sjnp#1#2#3{         {\it Sov. J. Nucl. Phys. }{\bf #1} (19#2) #3}
\def\yf#1#2#3{           {\it Yad. Fiz. }{\bf #1} (19#2) #3}
\def\zetf#1#2#3{         {\it Z. Eksp. Teor. Fiz. }{\bf #1} (19#2) #3}
\def\zetfpr#1#2#3{         {\it Z. Eksp. Teor. Fiz. Pisma. Red. }{\bf #1} (19#2) #3}
\def\jetp#1#2#3{         {\it JETP }{\bf #1} (19#2) #3}
\def\mpl#1#2#3{          {\it Mod. Phys. Lett. }{\bf #1} (19#2) #3}
\def\ufn#1#2#3{          {\it Usp. Fiz. Naut. }{\bf #1} (19#2) #3}
\def\sp#1#2#3{           {\it Sov. Phys.-Usp.}{\bf #1} (19#2) #3}
\def\ppnp#1#2#3{           {\it Prog. Part. Nucl. Phys. }{\bf #1} (19#2) #3}
\def\cnpp#1#2#3{           {\it Comm. Nucl. Part. Phys. }{\bf #1} (19#2) #3}
\def\ijmp#1#2#3{           {\it Int. J. Mod. Phys. }{\bf #1} (19#2) #3}
\def\ic#1#2#3{           {\it Investigaci\'on y Ciencia }{\bf #1} (19#2) #3}
\def\tp{these proceedings}
\def\pc{private communication}
\def\ip{in preparation}
\relax
\newcommand{\TeV}{\,{\rm TeV}}
\newcommand{\GeV}{\,{\rm GeV}}
\newcommand{\MeV}{\,{\rm MeV}}
\newcommand{\keV}{\,{\rm keV}}
\newcommand{\eV}{\,{\rm eV}}
\newcommand{\Tr}{{\rm Tr}\!}
\renewcommand{\arraystretch}{1.2}
\newcommand{\be}{\begin{equation}}
\newcommand{\ee}{\end{equation}}
\newcommand{\bea}{\begin{eqnarray}}
\newcommand{\eea}{\end{eqnarray}}
\newcommand{\ba}{\begin{array}}
\newcommand{\ea}{\end{array}}
\newcommand{\bmat}{\left(\ba}
\newcommand{\emat}{\ea\right)}
\newcommand{\refs}[1]{(\ref{#1})}
\newcommand{\ler}{\stackrel{\scriptstyle <}{\scriptstyle\sim}}
\newcommand{\ger}{\stackrel{\scriptstyle >}{\scriptstyle\sim}}
\newcommand{\lag}{\langle}
\newcommand{\rag}{\rangle}
\newcommand{\ns}{\normalsize}
\newcommand{\cm}{{\cal M}}
\newcommand{\gr}{m_{3/2}}
\newcommand{\p}{\partial}
\def\321{$SU(3)\times SU(2)\times U(1)$}
\def\tl{{\tilde{l}}}
\def\tL{{\tilde{L}}}
\def\bd{{\overline{d}}}
\def\tL{{\tilde{L}}}
\def\a{\alpha}
\def\b{\beta}
\def\g{\gamma}
\def\c{\chi}
\def\d{\delta}
\def\D{\Delta}
\def\db{{\overline{\delta}}}
\def\Db{{\overline{\Delta}}}
\def\e{\epsilon}
\def\l{\lambda}
\def\n{\nu}
\def\m{\mu}
\def\nt{{\tilde{\nu}}}
\def\p{\phi}
\def\P{\Phi}
\def\k{\kappa}
\def\x{\xi}
\def\r{\rho}
\def\s{\sigma}
\def\t{\tau}
\def\th{\theta}
\def\ne{\nu_e}
\def\nm{\nu_{\mu}}
\def\snui{\tilde{\nu_i}}
\def\la{{\makebox{\tiny{\bf loop}}}}
\renewcommand{\Huge}{\Large}
\renewcommand{\LARGE}{\Large}
\renewcommand{\Large}{\large}
\title{
\hfill hep-ph/0107204\\
\hfill TIFR/TH/01-26\\
Neutrino Anomalies in Gauge Mediated Model with \\
Trilinear R violation}

\author{ Anjan S. Joshipura~$^{a}$, Rishikesh D. Vaidya~$^{a}$ 
and Sudhir K. Vempati~$^{b}$ \\
{\ns\it $(a)$~ Theoretical Physics Group, Physical Research Laboratory,}\\
{\ns\it Navarangpura, Ahmedabad, 380 009, India.}\\
{\ns\it $(b)$~ Department of Theoretical Physics, Tata Institute of Fundamental
Research,}\\
{\ns\it Colaba, Mumbai 400 005, India.}}

\maketitle
\vskip 0.8 cm

\begin{center}
{\bf Abstract}
\end{center}

\begin{abstract}
The structure of neutrino masses and mixing resulting from trilinear
$R$ violating interactions is studied in the presence of the gauge mediated
supersymmetry breaking. Neutrino masses arise in this model at 
tree level through the RG-induced vacuum expectation values of the sneutrinos 
and also through direct contribution at 1-loop. The relative importance 
of these contributions is determined by the values of the strong and weak 
coupling constants.  In case of purely $\lambda'$ couplings, the tree 
contribution dominates over the 1-loop diagram. In this case, one 
simultaneously obtains atmospheric neutrino oscillations and
 quasi-vacuum oscillations of the solar neutrinos if all the $\l'$ 
couplings are assumed to be of similar magnitudes. If $R$ parity 
violation arises from the trilinear $\l$ couplings,  then the loop induced 
contribution dominates over the tree level.  One cannot simultaneously 
explain the solar and atmospheric deficit in this case if all the $\l$ 
couplings are of similar magnitude. This however becomes possible with 
hierarchical $\l$ and we give a specific example of this.
\end{abstract}

\newpage

\section{Introduction}

Variety of observations of the solar and atmospheric neutrinos 
have given important information on the possible structures of
neutrino masses and mixings. Based on these observations, one 
can infer that oscillations among three active neutrinos are likely
to be responsible for the observed features of the data. These
oscillations are characterized by two widely separated mass scales.
At least one mixing angle involved in oscillations of atmospheric
neutrinos is large \cite{skatm}. The detailed data on the day-night 
asymmetry and recoil energy spectrum of the solar neutrinos seem to 
favour the presence of one more large mixing angle \cite{sksolar}. 
Thus the neutrino mass spectrum seems to be characterized by hierarchical 
(mass)$^2$ differences and by two large and one small mixing angles,
with the small mixing angle demanded by the CHOOZ experiment \cite{choose}.
Many mechanisms have been advanced to understand these features of 
the neutrino spectrum \cite{revs}.  One of these is provided by 
supersymmetric theory which contains several features to make it attractive 
for the description of the neutrino spectrum.  The lepton number violation 
needed to understand neutrino masses is in-built in this theory through 
the presence of the  $R$ parity violating couplings \cite{hall}. 
Moreover, it is possible to understand the hierarchical neutrino masses 
and large mixing among them  within this framework without fine tuning 
of parameters or without postulating ad hoc textures for the neutrino mass 
matrices \cite{asjbabu,asjskv}. 

The supersymmetrized version of the standard model contains the following 
lepton number violating couplings:

\be
\label{doubleul}
W_{\not{L}} = \e_i L_i H_2 + \l_{ijk} L_i L_j e_k^c + \l'_{ijk} L_i Q_j d_k^c,
\ee
where $L, Q, H_2$ represent the leptonic, quark and one of the  Higgs
doublets (up-type) respectively and $e^c$, $d^c$ represent the leptonic 
and down quark singlets.  Each of these couplings is a potential source 
for neutrino masses.
There have been detailed studies of the effects of these couplings
on neutrino masses under different assumptions 
\cite{asjbabu,asjskv,bilinear_univ,bilinear_nouniv,romao,kaplan}. 
We briefly recapitulate the relevant gross  features of these studies and 
motivate additional work that we are going to present in this paper. 

The most studied effect is that of the three bilinear mass parameters
$\e_i$, particularly in the context of the supersymmetry breaking with
universal boundary conditions at a high scale 
\cite{asjbabu,asjskv,bilinear_univ,romao,kaplan}. 
This formalism provides a nice way of understanding suppression in 
neutrino mass $m_\nu$
compared to the
weak scale. The mixing among neutrinos is largely controlled in this
approach by the ratios
of the parameters $\e_i$. This can be approximately
described by the following matrix for a large range in parameter space 
of the minimal supersymmetric standard model with radiative $SU(2)\times
U(1)$ breaking \cite{asjbabu,asjskv,kaplan}:
\be
\label{bilinearmix}
U = \bmat{ccc}
c_1 & s_1 c_2 & s_1 s_2\\
- s_1 & c_1 c_2 & c_1 s_2\\
0 & - s_2 & c_2 \emat , 
\ee

where 
\be
s_1 = {\e_1 \over  \sqrt{\e_1^2 + \e_2^2}} \;\;\;;\;\;\;\; s_2 = 
\sqrt{{\e_1^2 + \e_2^2 \over \e_1^2 + \e_2^2 + \e_3^2 }}.
\ee

The above matrix can  nicely reproduce the small angle MSW
solution together with the atmospheric neutrino anomaly. However,
due to the specific structure, one cannot have two large mixing angles
keeping at the same time $|U_{e3}|\leq 0.1$ as required from the negative
results of CHOOZ.
Thus purely bilinear $R$ violation with universal boundary conditions
cannot account for the observed features with two large mixing angles.
This may be remedied by 
not insisting upon the universal boundary conditions \cite{romao}
 or by adding right handed neutrinos \cite{rhneut}.

In contrast to the bilinear case, the presence of trilinear interactions
can allow  two large angles without conflicting with the CHOOZ result.
There have been several studies to
determine possible set of trilinear couplings which can reproduce the
observed features of neutrino masses and mixing \cite{abada}.
It is not surprising that one could `fit' the neutrino spectrum in these
cases due to very large number of trilinear couplings. But it was realized
\cite{drees,sneutrino}
that gross features of the neutrino spectrum can be understood without
making specific assumptions on the trilinear couplings
other than requiring them to be  similar in magnitude.
This makes the $R$ violation with trilinear interaction `predictive'
in spite of the presence of very large number of couplings. 

In addition to the $R$ violating parameters, the neutrino spectrum in these
models also depends upon the nature of the supersymmetry breaking. This spectrum
has been studied in the standard supergravity case with bilinear 
as well as trilinear
couplings and in the case of gauge mediated supersymmetry breaking when
$R$ violation is only through the bilinear terms in eq.(\ref{doubleul}). 
The supersymmetry breaking generically introduces two different types of 
contributions to neutrino masses.  The presence of terms linear in the 
sneutrino field in the scalar potential induces a vacuum expectation value 
({\it vev}) for the former which mix neutrinos with neutralinos
and lead to neutrino masses. In addition to this `tree level' contribution, 
the trilinear terms in the superpotential also lead to neutrino masses 
at the 1-loop level. Both the types of contributions are present in models with 
purely bilinear or purely trilinear terms in the superpotential at a high scale.
In case of the trilinear couplings, the running of the couplings to lower scale 
generates in the scalar potential couplings linear in the sneutrino {\it vev} 
and lead to a tree level contribution which is often neglected in the 
literature.

The relative importance of the tree level and loop induced contributions to 
neutrino masses in case of trilinear interactions was studied in 
\cite{sneutrino,carlos} in the context of the standard supergravity (mSUGRA)
 models with universal boundary conditions at a high scale. It was concluded 
that the tree level  contribution dominates over the loop for large ranges
in the parameters of the model. This results 
\cite{sneutrino} in the following hierarchy in neutrino masses if all the 
trilinear $\l'$ couplings are assumed to be similar in magnitudes:   
\be
{m_{\nu_2} \over m_{\nu_3}} \approx {m_\la \over m_0 + m_\la} {m_s \over m_b} ~.
\ee 
The parameters $m_\la$ and $m_0$ characterize the strength 
of the 1-loop  and the RG-induced tree level contributions respectively and are
determined by soft SUSY breaking parameters. $m_s$ and $m_b$ denote
the strange quark and the bottom quark masses respectively. 
It is possible to obtain
the vacuum or the MSW solution (large angle) to solar neutrino 
problem in this context by choosing the SUSY parameters in appropriate range\cite{sneutrino}. 

An attractive alternative to the standard supergravity induced SUSY breaking 
is provided by the gauge mediated SUSY breaking \cite{dine}. 
 Neutrino mass spectrum has been studied in gauge mediated
models with trilinear R-violation by Choi {\it et al.} in
\cite{chun}. Their study has been confined to non-minimal models 
of this category.
The minimal model in this category called the Minimal Messenger
Model(MMM)  \cite{mmm}
 has only two free parameters and is more predictive than the standard 
SUGRA based models and the models studies in \cite{chun}. The two 
free parameters of the model determine all the soft terms at the high scale
$\sim $ a few hundred TeV, where SUSY breaking occurs. 
Thus this model implies very constrained spectrum 
for neutrino masses.  This constrained spectrum has been shown 
\cite{asjskv} to be inadequate for simultaneous solution of the solar 
and atmospheric neutrino anomalies in case of purely bilinear $R$ violation.
 In this work, we wish to study neutrino masses in the minimal messenger model
 in the presence of purely trilinear $R$ violation. This would mean that both 
the scale at which SUSY breaking occurs as well as the boundary conditions 
at the high scale would be sufficiently different from the mSUGRA scenario 
which has been studied in \cite{sneutrino}. We have studied the neutrino mass
spectrum in the MMM for two separate cases, namely purely $\l'$ couplings
and purely $\l$ couplings. Such a choice has been made for simplicity.  
The $\l'$ couplings with comparable magnitude are argued to
describe neutrino spectrum well. In contrast, we find that if all the $\l$
couplings are of similar strength, then one cannot describe the neutrino 
spectrum well and one needs to postulate somewhat inverse hierarchy among 
them.  We give a specific example with hierarchical $\l$ which reproduces 
the observed features of the neutrino spectrum.  

Within the Minimal Messenger Model, the soft supersymmetry breaking terms 
are decided by  the gauge quantum numbers of the fields.  As we will 
demonstrate later, this significantly alters the hierarchy within 
the neutrino mass states.  In particular, we find that  the $m_0$ 
dominates over $m_\la$ in case of the $\l'$ couplings but the situation 
is reversed when the $R$ violation occurs through $\l$ couplings. 
This feature is characteristic of the gauge mediated scenario and is 
quiet distinct from all the earlier studies.

We discuss the basic formalism in the next section which also contains
analysis of the effect of the trilinear $\l'$ couplings. The third section
has detailed study of the $\l$ couplings and we end with a discussion in the 
last section. 

\section{Formalism}

We consider the following trilinear interaction in this section:
\be
W_{\not{R}_p} = \l'_{ijk} L_i Q_j d_k^c ~,
\ee
where $i,j,k$ are generation indices. 
In spite of very large number of these couplings, one could determine the
neutrino masses and mixing in terms of small number of parameters if one
assumes that all the trilinear couplings are similar in magnitude. The
basic formalism was developed in \cite{sneutrino} and we recapitulate
here the relevant parts.

The neutrinos obtain their masses from two different contributions 
in this case. The $\l'$ couplings generate radiative masses through
exchange of the down squarks at the 1-loop level. In addition to this,
the trilinear interactions also radiatively generate  soft SUSY breaking
terms which are linear in the sneutrino fields. These terms lead to
additional contribution to neutrino masses which can dominate over the
the first contribution. The second contribution follows from the RG improved
scalar potential \cite{sneutrino,carlos}:
\bea \label{soft}
V_{scalar}&=& m_{\tilde{\n}_i}^2 \mid \tilde{\n}_i \mid ^2 + m_{H_1}^2
\mid H_1^0 \mid^2 + m_{H_2}^2 \mid H_2^0 \mid^2  + \left[ m_{\n_i H_1}^2
\tilde{\n}^{\star}_i H^0_1 \right. \nonumber \\
& & \left. - \m\;B_\m H_1^0 H_2^0 - B_{\e_i} \tilde{\n}_{i} H_2^0  + H.c
\right]
+ {1 \over 8} (g_1^2 + g_2 ^2) (\mid H_1^0 \mid  ^2 -
 \mid H_2^0 \mid ^2)^2 + .... \;\;\; ,
\eea
where we have used the standard notation for the SM fields and their masses, 
 with $B_{\e_i}$ and $m_{\n_i H_1}^2$  
representing the bilinear lepton number violating soft terms. 
Minimization of the above potential leads to the sneutrino {\it vevs}:
\be
\label{omega}
< \tilde{\n_i} >~~   = {B_{\e_i} v_2 - m^2_{\n_i H_1} v_1 \over m_{L_i}^2
+ {1 \over 2} M_Z^2~ \cos 2 \beta }\;\;\;,
\ee
where $v_1$ and $v_2$ stand for the {\it vevs} of the Higgs fields $H_1^0$ and
$H_2^0$ respectively \footnote{These sneutrino {\it vevs} are derived from
the tree level scalar potential. Corrections from the one-loop effective 
potential can significantly shift these naive tree level values 
\cite{gamberini}. For neutrino phenomenology these corrections would
be important in regions in the parameter space where two contributions to
the sneutrino {\it vev} cancel each other \cite{chun1loop}. 
Such regions are not encountered in MMM parameter space in which we 
are interested. Moreover, we are approximately including the effect of 1-loop
corrections by dynamically choosing soft parameters at appropriate scale in 
the manner discussed in Refs.\cite{gamberini,borzu}.}. 

These {\it vevs} vanish at a high scale since we are
assuming only trilinear $L$ violating interactions. They however 
get generated at the weak scale. The magnitudes of the parameters 
$B_{\e_i}$ and $m_{\n_i H_1}^2$ and hence the sneutrino {\it vevs} at the weak
scale are determined by
solving the renormalization group (RG) equations satisfied by them. These
RG equations are presented in Appendix I. 
The general solution of these equations can be
parameterized as

\bea
\label{sol}
B_{\e_i}&=& \l'_{ipp} h^D_{p} \kappa_{ip}\;, \nonumber \\
m_{\n_i H_1}^2&=&\l'_{ipp} h^D_{p} \kappa'_{ip}\;,
\eea
where $\k,\k'$ are dependent on the soft terms appearing in the RHS of
the respective RG equations and $h^D$ are down type quark yukawa.

The sneutrino {\it vevs} break $R$ parity and lead to mixing of neutrinos
with
neutralinos. This in turn leads to neutrino masses. For small sneutrino
{\it vevs}, the neutrino mass matrix follows from the seesaw 
approximation and is given by \cite{asjmarek}:

\bea
\label{mnot}
{\cal M}^0_{ij}&=&{\m (cg^2 + g'^2) ~ < \tilde{\n_i} >~ < \tilde{\n_j} >
\over 2 ( -c \m M_2 + 2~ M_W^2 c_\b s_\b (c + \tan \theta_W^2 ))} \;\; ,
\eea

where $c = {5 g'^2/3  g^2 } \sim 0.5$, $M_W$ is the $W$ boson mass,
$\m$ is the mass term for Higgs/Higgsino, $M_2$ is mass term for a
gaugino, 
and the Weinberg angle is represented by $\theta_W$. 
Assuming generation independence of the terms $\kappa,\kappa'$  which
was found to be a very good approximation in \cite{sneutrino}, we can
rewrite the above mass matrix as  
\be
\label{mnotp}
{\cal M}^0_{ij} \equiv m_0 ~\l'_{ipp}~ h^D_p ~\l'_{jmm}~ h^D_m \;\; ,
\ee
where the parameters $p$ and $m$ are summed over the three generations 
and  $m_0$ now contains the dependence of the tree level mass on 
the soft SUSY breaking parameters. Only one neutrino attains mass
through this mechanism. The other neutrinos attain mass at the 
1-loop level.  The complete 1-loop structure of the neutrino masses
has been discussed in \cite{chun1loop}. In the present case, the most 
dominant contributions are from diagrams having $\l'$ couplings at both
the vertices. 
The mass matrix generated  by these diagrams is given by 
\be
\label{mloop}
{\cal M}^{l}_{ij} = {3 \over 16 \pi^2 } \l'_{ilk} \l'_{jkl} ~ v_1~ h^D_{k} ~\sin
\phi_l\; \cos \phi_l~ \ln {M_{2l}^2 \over M_{1l}^2}\;\;\;.
\ee

In the above, $\sin\phi_l\; \cos\phi_l$ determines the mixing of the
 squark-antisquark pairs and $M_{1l}^2$ and $M_{2l}^2$ represent
the eigenvalues of the standard 2$\times$ 2 mass matrix of the
down squark system. The indices $l$ and $k$ are summed over.  
As the mixing $\sin\phi_l\; \cos\phi_l$ is proportional to $h^D_l$, we
rewrite the 1-loop contribution as, 

\be
{\cal M}^l_{ij} = m_\la ~ \l'_{ilk}~ \l'_{jkl}~ h_{k}^D~
h_{l}^D, 
\ee

\noindent where $m_\la$ is independent of the $R$ violation and solely
depends on the MSSM parameters.

The total neutrino mass matrix is given by,
\be
{\cal M}^\n = {\cal M}^0 + {\cal M}^l
\ee
which can be rewritten in the following form when 
${\cal O}(h_1^{D^2}, h_2^{D^2})$ terms are neglected:

\be
\label{mtotal2}
{\cal M}_{ij}^\n \approx (m_0 + m_\la )~ a_i a_j + m_\la~ h_{2}^D~
h_{3}^D~ A_{ij} 
\ee
where $a_i = \l'_{ipp} h^D_p$ ($p$ summed over generations) and 
\be
\label{aij}
A_{ij} = \l'_{i23} \l'_{j32} + \l'_{i32} \l'_{j23} - \l'_{i22} \l'_{j33} -
 \l'_{i33} \l'_{j22}\;\; .
\ee

To derive the eigenvalues of the total matrix ${\cal M}^\nu$, we recognise
that a) The first matrix on the RHS of eq.(\ref{mtotal2}) has only one
non-zero eigenvalue ;  b) The dominant terms in the total matrix 
${\cal M^{\nu}}$ of ${\cal O}(h_3^{D~2})$ are present only in the 
first matrix . Moreover, as we will show below $m_\la \ll m_0$ 
in the Minimal Messenger Model in the purely $\l'$ case. Hence approximate
eigenvalues can be derived up to ${\cal O}({ m_s m_\la \over m_b ( m_0 +
m_\la )} )$, neglecting the high order corrections.  The detailed derivation 
of the eigenvalues and the mixing matrix has been presented in \cite{sneutrino}.
 These eigenvalues are given as, 

\bea
\label{evalues}
m_{\n_1}&\sim & m_\la h_2^D  h_3^D ~\d_1 \nonumber \\
m_{\n_2}&\sim&  m_\la h_2^D  h_3^D ~\d_2 \nonumber \\
m_{\n_3}&\sim& (m_0  + m_\la) \sum_i^3 a_i^2 \nonumber \\
&\sim& (m_0  + m_\la)  {h_3^D}^2 \sum_i^3 {\l^\prime _{i33}}^2
\eea

where

\bea \label{deltas}
\d_1 &= &  ~(c_1^2 ~A'_{11} - 2 c_1 s_1 A'_{12} +
 s_1^2 A'_{22}) \;\;,\nonumber \\
\d_2 &= &   ~(s_1^2 ~A'_{11} + 2 c_1 s_1 A'_{12} +
 c_1^2 A'_{22})\;\; .
\eea

The entries $A'_{ij}$  are the elements of the matrix 
$A' = U_{\l'}^T A U_{\l'}$ where

\be
\label{treemix}
U_{\l'} = \left(
\ba{ccc}
c_2&s_2 c_3&s_2 s_3\\
-s_2&c_2 c_3&c_2 s_3\\
0&-s_3&c_3\\ \ea \right)\;\; ,
\ee

with

\be 
s_2={a_1 \over \sqrt{a_1^2 + a_2^2}}~\;\;;\;\;s_3= \sqrt{{a_1^2 + a_2^2
\over a_1^2 + a_2^2 + a_3^2}}\;\; .
\ee

The total mixing is given as \cite{sneutrino},
\bea \label{mix}
K'&=&U_{\l'}~U'_{\l'}\nonumber \\
&=&\left( \ba{ccc}
c_1 c_2 - s_1 s_2 c_3&  s_1 c_2 + c_1 s_2 c_3 & s_2 s_3 \\
-s_2 c_1 - s_1 c_2 c_3&- s_1 s_2 + c_1 c_2 c_3 & c_2 s_3 \\
s_1 s_3 & -s_3 c_1 & c_3 \ea \right),
\eea

where the 1-2 mixing angle $\theta_1$ is given by,

\be
\label{12}
\tan 2 \th_1 = { 2 A'_{12} \over A'_{22} - A'_{11}}\;.
\ee 

From eq.(\ref{aij}), we see that in the limit of exact degeneracy of the
$\l'$ couplings, the parameters $A_{ij}$ would be zero. In this case,
only one neutrino becomes massive in spite of the inclusion of the 
loop corrections. The 1-2 mixing also remains undetermined in this
case. There is no reason {\it a priori} for the exact equality of $\l'$
and non-zero but similar value for these parameters determine 
the 1-2 mixing to be large (see eq.(\ref{12})) and also generates mass
for the other two neutrinos.

\subsection{MMM and neutrino anomalies}

The parameters $m_\la$ and $m_0$ appearing in  eq.(\ref{evalues}) are 
independent of the details of the $R$ violation and get determined by the 
soft SUSY breaking terms. We assume throughout 
that SUSY breaking is mediated by the standard gauge interactions 
\cite{dine}.  We work in the so-called minimal messenger model \cite{mmm}.
It is characterized by a messenger 
sector with a pair of superfields which transform vector-like under a 
gauge group chosen to be $SU(5)$ for unification purposes. SUSY breaking is
characterized by a singlet chiral superfield whose  scalar and the
auxiliary components  acquire {\it vevs} breaking supersymmetry. This breaking
is communicated to the visible sector by loop diagrams. The gauginos acquire 
masses at the 1-loop level which are given as,

\be
\label{gmass}
M_i (X) = \tilde{\alpha}_i(X)~ \Lambda~ g(x)~,
\ee
where $X = \lambda <S>$ is the supersymmetric mass of the scalar and fermionic
components of the singlet superfield, 
$\Lambda $ is the ratio ${F_S \over <S>}$, $F_S$ 
being the {\it vev} of the auxiliary field of the singlet. The parameter $x$ is
defined as ${\Lambda \over X}$. The scalars acquire masses at the two 
loop level. They are given by
\be
\label{smass}
m_i^2 = 2 \Lambda^2 \left( C^i_3 \tilde{\alpha}_3^2 (X) + C^i_2 
\tilde{\alpha}_2^2 (X) + {3 \over 5} Y_i^2 \tilde{\alpha}_1^2 (X) \right) f(x)~.
\ee
where $C_3, C_2$ are the quadratic casimirs of the gauge groups 
$SU(3)$ and $SU(2)$ respectively and $Y_i$ being the hypercharge of
the scalar field $i$ , with $i= \{ Q_j,~D_j,~U_j,~L_j,~E_j,~H_1,~H_2 \}$,
where $j = 1,2,3$ is the generation index. 
The functions $f(x),g(x)$ are given in \cite{martin} and the dependence
of the soft masses on these functions is minimal. In the present analysis,
we follow \cite{borzu} and choose $x = {1 \over 2}$. Since the dependence of the 
soft masses on $x$ is minimal, a different choice of $x$ would not significantly
modify the results presented here.  

The major feature characterizing the model is the absence of $A$ -terms
and the $B$ terms in the soft potential  at the
scale $X$.
\be
\label{highb}
A(X) = 0 \;\;,\;\; B(X) = 0~.
\ee
Thus, the entire soft spectrum of
this model gets essentially determined by one parameter $\Lambda$. The
parameters tan$\beta$ and  $\mu$ are fixed at the weak scale by requiring 
the breaking of the $SU(2)\times U(1)$ symmetry. The relevant equations
following from the tree level potential are given by
\bea
\label{mini}
\sin 2 \beta &=& {2 B_\m \m  \over m_{H_1}^2 + m_{H_2}^2 + 2 \m^2 } \nonumber \\
\m^2 &=& { m_{H_1 }^2 - m_{H _2}^2 \tan^2 \b \over \tan^2 \beta - 1} - 
{1 \over 2} M_Z^2 
\eea
where all the parameters on the RHS of the above equations are evolved
to the weak scale using the MSSM RGE. Because of  the boundary condition eq.(\ref{highb}),
the value of the $B$ parameter at the weak scale remains small. This pushes
tan$\beta$ to very large values in this model \cite{borzu,rattazzi}. 
The above equations are
strictly valid in case of the MSSM and $R$ violating trilinear couplings 
can give corrections to these. The smallness of neutrino masses however
require very tiny trilinear couplings of O$(10^{-4} - 10^{-5})$ which would
contribute negligible to eqs.(\ref{mini}). We thus continue 
to use eqs.(\ref{mini}). 

The neutrino masses in eq.(\ref{evalues}) are strongly hierarchical 
in the limit  $m_\la\ll m_0$.  Specifically, one obtains 
from eq.(\ref{evalues}),
\be
\label{ratio}
{m_{\n_2} \over m_{\n_3}} \sim {m_\la \over m_0} {m_s \over m_b} 
\left( \delta_2 \over \sum_i (\l'_{i33})^2 \right)
\ee
where $m_s$ and $m_b$ represent the strange and the bottom quark 
masses. For all the $\l'$ of similar magnitudes, this ratio is 
completely determined by the parameter $\Lambda$. 

We can determine the above ratio by exactly solving the RG equations, 
(\ref{lprge}) . 
Before doing this, it is instructive to study the approximate 
expressions obtained when one neglects the $Q^2$ dependence of the parameters
appearing on the RHS of the RGE. In other words, we neglect
the effect of running of the soft masses (from high scale, $X$ to weak 
scale $M_Z$) appearing in the expressions of the RGE as well as for 
neutrino masses. Instead, we take them to be their high scale values given 
by eqs.(\ref{smass}).  Noting that more dominant contribution to 
sneutrino {\it vev} comes from the $m^2_{\nu_i H_1}$ term in 
eq.(\ref{omega}) and integrating the corresponding RGE for $m^2_{\nu_i H_1}$ 
in the above approximation, one finds in this simplifying case :

\be
\label{apm0mmm}
m_o \sim \left({2 \cos\b\over 3 \pi^2} \right)^2 
{ M_W^2\over \Lambda} ~~
{ \tilde{\alpha}_3^4(X) \over \tilde{\alpha}_2^5(X)}~~
 \left(\ln { X^2 \over M_Z^2} \right)^2
\ee
The $m_\la$ defined in eq.(\ref{mloop}) has the following approximate form
in the same approximation as above where we neglect the running of the soft
masses. 
\be
\label{apmlmmm}
m_\la \sim \left( { v^2 \mu \over \Lambda^2 } \right) 
{\cos \beta~\sin\b \over 8 ~\pi^2~\tilde{\alpha}_2 ^2(X)}
\ee

Eq.(\ref{apm0mmm}) clearly demonstrates that the often neglected 
\cite{abada,drees} RG induced contribution dominates over the loop 
contribution in the present case of MMM just as in the case of the 
supergravity induced breaking \cite{sneutrino,hera1}.
One would have naively thought that this will not be the case in gauge
mediated model since the running of masses in this case (signified by 
$t \equiv \ln({X^2 \over M_Z^2} ) \sim 11$ in eq.(\ref{apm0mmm})) 
is over much smaller range than in the supergravity case where 
$t\equiv \ln({M_{GUT}^2 \over M_Z^2}) \sim 66.$ But smallness 
of $t$ in MMM is compensated by
the largeness of the ratio ${m_{\tilde{Q}}^2 \over m_{\tilde{L}}^2}$
(signified by the factor) $ \tilde{\alpha}_3^4 \over \tilde{\alpha}_2^4$.
As a result of which the value of $m_0$ here can be comparable to the
corresponding value \cite{sneutrino} in supergravity case. 
From the expressions we see that dependence on the $\Lambda$ is more severe
for the 1-loop mass, $m_\la$ compared to the tree level contribution,
$m_0$. However, the $\mu$ parameter in the numerator increases approximately 
linearly with $\Lambda$ \cite{borzu}. This makes the $\Lambda$ 
dependence of the both the contributions essentially the same. The ratio 
${m_\la \over m_0}$ following from eqs.(\ref{apm0mmm},\ref{apmlmmm})
is given as:

\be
\label{apratiolamp}
{m_\la \over m_0} \sim \left({\pi^2\over 3 }\right) 
\left({ \m~ \tan \beta~ v^2\over M_W^2~\Lambda~ t^2 } \right)
\left({ \tilde{\alpha}_2^3(X) \over \tilde{\alpha}_3^4 (X)} \right)
\ee
From the above we see that the dependence of the ratio on $\Lambda$ is 
essentially determined by the way the $\mu$ parameter scales with respect
to $\Lambda$. This leads to a very mild dependence of the ratio on $\Lambda$.
Such $\Lambda$ `independence' has also been seen in the case 
of bilinear R-violating models in MMM \cite{asjskv}. 
For $\Lambda = 100$ TeV, $ t = 2~ \ln \left({X^2 \over M_Z^2} \right)
 = 10.6 $, $\tan \beta = 46$ and $\m = 400$ GeV, from the above we see
that the ratio is  $ 0.39$.  
From eqs.(\ref{apm0mmm},\ref{apmlmmm}) we see that the typical order
of magnitude for the ratio of the mass eigenvalues, eq.(\ref{ratio}) 
is :
\be
\label{nratlamp}
{m_{\n_2} \over m_{\n_3} } \sim 10^{-3}
\ee
However the above expressions are approximate. 
We have determined this ratio exactly  by solving the relevant RG equations
numerically.
For this, we closely follow the work of \cite{borzu} where
two loop RGE for the $B$ parameter were used for fixing the sign of $\m$ 
parameter at the weak scale. As mentioned above, we have
chosen  $x = {1 \over 2}$. This fixes $ X = 2 \Lambda$. Following
\cite{borzu}, the decoupling scale `$Q_0$' is defined to be the 
geometrical mean of  $m_Q^2(X), m_U^2(X)$. Such a choice of $Q_0$
would be approximately equivalent to adding corrections due to the
complete 1-loop effective potential. The gauge couplings and the
Yukawa couplings are evolved from $M_Z$ to $Q_0$ using the Standard
Model (SM) beta-functions and using MSSM beta-functions from $Q_0$ to 
the high scale, $X$.  All the soft masses and soft parameters
defined at $X$ are evolved from $X$ to the decoupling scale $Q_0$. 
The ratio ${m_{\nu_2} \over m_{\nu_3}}$, determined following the above 
procedure is plotted in  Figure 1 for $\Lambda$ varying from $(50 - 150)$ TeV. 

From the figure we see that the ratio of the eigenvalues, 
\be
\label{calratiolamp}
{m_{\n_2} \over m_{\n_3}} \approx  (1-2) \times 10^{-4} 
\left( \delta_2 \over \sum_i (\l'_{i33})^2 \right)
\ee 
is typically around the expected value, eq.(\ref{nratlamp}). While the ratio shown in Fig. 1
is completely fixed by the value of $\Lambda$, the neutrino mass ratio is uncertain by
a number of ${\cal O}(1)$ which is related to the trilinear parameters.

We now turn to discussing feasibility of the model for the simultaneous description of the 
solar and atmospheric data.  The two generation analysis of each of these experiments
constrain the value of the relevant $(mass)^2$ difference and mixing angle. The allowed
values for the parameters \cite{gonzalez,oyama,sno} are displayed below:
\vskip 0.6 cm
\begin{tabular}{|c|c|c|c|}
\hline
Anomaly&Solution&$\Delta m^2 (eV^2)$& $ \tan^2 \th$\\[3pt]
\hline
Solar& ~~~MSW-SMA&$(2-10) \times 10^{-6}$&$(1-20) \times 10^{-4}$\\[3pt]
& MSW-LMA&$~~~ (2-80) \times 10^{-5}$& 0.2~-~ 4. \\[3pt]
& LOW-QVO& ~~~$1 \times~10^{-10}- 4 \times 10^{-7}$&0.1 -8.  \\[3pt]
& Vacuum (Just-So)& ~~~$(4-12) \times~10^{-12}$&0.1 -7.  \\[3pt]
\hline
\hline
Anomaly&&$\Delta m^2 (eV^2)$& $ \sin^2 2\th$\\[3pt]
\hline
Atmosphere&&~~~$(1-8) \times~10^{-3}$&0.83 - 1.\\[3pt]
\hline
\end{tabular}
\vskip 0.6 cm

At present, global analysis including the recent results from the 
day/night recoil electron energy  spectrum and charged current rates
from SNO of the solar neutrinos from
super-Kamiokande \cite{sksolar} favours the MSW-LMA solution
but  the other solutions are also allowed \cite{gonzalez,sno}. It follows from 
eq.(\ref{calratiolamp})
and the table that the most natural solution for the solar neutrino problem
is through the vacuum oscillations but quasi-vacuum solution can also be obtained,
if the $\l'$ dependent factor in eq.(\ref{calratiolamp}) 
is somewhat large (e.g. $\sim 5$) instead of exactly being one. 
The most preferred LMA solution cannot however be obtained.

The mixing among neutrinos is essentially controlled by ratios of
trilinear couplings. This mixing is given by eq.(\ref{mix}). It is seen
from this equation that a choice of angles $s_{1,2,3}$ is possible 
which reproduces two large and one small mixing angles as required by the
present data. As an example, consider the choice  
\be
c_1 \;\;= c_3 \;\;=s_1 \;\;=s_3 \;\;= {1 \over \sqrt{2}}\;\;;s_2\sim .13~.
\ee
This gives 
\bea
\sin^2 2 \th_{A} &\equiv& 4 K^{'~2}_{\m 3} (1- K^{'~2}_{\mu 3}) \approx 0.99~, \nonumber \\
\sin^2 2 \th_{S}&\equiv& 4 K^{'~2}_{e1}~K^{'~2}_{e2}\approx 0.95~, 
\nonumber \\ 
K'_{e3}&=& 0.09 ~.
\eea
This choice reproduces the required mixing angles and also satisfies 
the CHOOZ constraint. 

\section{ Models with $\l_{ijk}$} 

In this section, we discuss the structure of neutrino masses and mixing
in the presence of only trilinear $\l$ couplings. The lepton number
violating part of the superpotential is given as,
\be
\label{superlamb}
W_{\not{L}} = \l_{ijk} L_i L_j e_k^c~.
\ee
There are two basic changes here compared to the last section. Firstly,
the $\l_{ijk}$
are antisymmetric in the first two indices restricting their total number to nine.
This strongly restricts neutrino mass structure and one does not get 
phenomenologically consistent spectrum when all the trilinear couplings 
are assumed to be similar in magnitudes. Secondly, unlike in the $\l'$ case, 
the loop induced contribution dominates over the tree level SUSY breaking 
in the minimal messenger model.  Such dominant loop
contribution has been earlier seen in some particular regions of the 
parameter space in the mSUGRA framework \cite{sneutrino,chun}.
Here this dominance follows generically.

We present here a formalism which is similar to the spirit of the 
previous section in order to understand the basic features of the
neutrino mass matrix . As before, the RG improved effective
soft potential contains terms $B_{\e_i}$ and $m^2_{\n_i H_1}$ which break
the lepton number. These terms are generated due to the presence of $\l$
couplings in the superpotential. At the weak scale, the magnitude of these 
depends on the respective RGE, which we have presented in Appendix I. The
solutions of  eqs.(\ref{lrge}) can be written as,
\bea
B_{\e_i}&=& \l_{ipp}h^E_{pp} \tilde{\k}_{ip}, \\~
m^2_{\n_i H_1}&=& \l_{ipp} h^E_{pp} \tilde{\k}'_{ip}~,
\eea
where $(i \neq j)$ due to the anti-symmetric nature of the $\l$ couplings
and $\tilde{\k}$ and $\tilde{\k}'$ represent the dependence on the soft
masses in the RGE and $h^E$ are the charged lepton yukawa.  Following
similar arguments as in Section II, 
the presence of these terms in the scalar potential would lead to a 
tree level neutrino mass matrix of the following form:
\be
{\cal M}^{0}_{ij} = m_0 b_i b_j~,
\ee
where $m_0$ contains the dependence on the soft terms and $b_i$ are
given as, $b_i = \l_{ipp} h^E_{pp}~~ (i \neq p) $. In addition to
the tree level mass, the presence of $\l$ couplings also gives rise
to contributions at the 1-loop level.  Assuming only canonical 1-loop 
contributions to be the most dominant contributions the 1-loop level 
mass matrix has the form:
\be
\label{mloopl}
{\cal M}^{l}_{ij} =  {1 \over 16 \pi^2 } \l_{ilk} \l_{jkl} ~ v_1~ h^E_{k} ~\sin
\phi_l\; \cos \phi_l~ \ln {M_{2l}^2 \over M_{1l}^2},
\ee 

where $ \sin \phi_l \cos \phi_l$ and $M_{1l}, M_{2l}$ represent the mixing
and the eigenvalues respectively of the standard $2\times2$ stau slepton
mass matrix. Following the previous section, we rewrite the above as, 

\be
\label{mloopla}
{\cal M}^{l}_{ij} =  m_\la~\l_{ilk} \l_{jkl} h^E_l h^E_k
\ee

In writing the above, we have implicitly assumed the anti-symmetric nature 
of the couplings. The total neutrino mass matrix is given as,
\be
{\cal M} = {\cal M}^0 + {\cal M}^l
\ee
The above can be rewritten in the form:
\be
\label{altrue}
{\cal M}_{ij} = (m_0 + m_\la) b_i b_j + m_\la h_3^E h_2^E B_{ij} +
{\cal  O} ({h_2^E}^2, h_2^E h_1^E),
\ee
where we have neglected ${\cal O} ({h_2^E}^2,h_2^E h_1^E)$ contributions to the
mass matrix.  The matrix $B$ is given as

\be
B = \bmat{ccc}
\l_{132} \l_{123} - \l_{133} \l_{122} & \l_{123} \l_{232} - \l_{122} \l_{233}
& \l_{132} \l_{323} - \l_{133} \l_{322} \\
\l_{123} \l_{232} - \l_{233} \l_{122} & 0 & 0 \\
\l_{132} \l_{323} - \l_{133} \l_{322} & 0 & 0 \emat
\ee

We diagonalise the total matrix ${\cal M}$ in the same manner as for the
$\l'$ case and as described in \cite{sneutrino}. However we do not make 
any assumption on the relative magnitude of $m_0$ and $m_\la$. 
The approximate eigenvalues correct up to ${\cal O}(h_2^{E~2})$ 
are derived as, 

\bea
\label{levalues}
m_{\nu_1}& \approx & m_\la h_3^E h_2^E \eta_1 + {\cal O}(h^{E~2}_2) \nonumber \\
m_{\nu_2}& \approx & m_\la h_3^E h_2^E \eta_2 + {\cal O}(h^{E~2}_2) \nonumber \\
m_{\nu_3}& \approx & (m_0 + m_\la) \sum_i^3 b_i^2 \nonumber  \\
 & \approx & (m_0 + m_\la) h_3^{E~2} \eta_3 + 
{\cal O}(h^E_3 h^E_2) 
\eea 

The parameters $\eta_1,\eta_2, \eta_3 $ are given by,
\bea \label{etas}
\eta_1 &= &  ~(c_1^2 ~B'_{11} - 2 c_1 s_1 B'_{12} +
 s_1^2 B'_{22}) \nonumber  \\
\eta_2 &= &   ~(s_1^2 ~B'_{11} + 2 c_1 s_1 B'_{12} +
 c_1^2 B'_{22}) \nonumber  \\
\eta_3 &= & \sum_{i=1,2} \l_{i33}^2 
\eea

The parameters $B'_{ij}$ are elements of the matrix, 
$B' = U_{\l}^T B U_{\l}$ where, $U_{\l}$ has the same form as
$U_{\l'}$ in eq.(\ref{treemix}) with the angles now given by,

\be \label{s2s3}
s_2 = {b_1 \over \sqrt{b_1^2 + b_2^2}} \;\;,\;\; s_3 = \sqrt{{ b_1^2 + b_2^2 
\over b_1^2 + b_2^2 + b_3^2 }}
\ee

The total mixing matrix is given as in eq.(\ref{mix}) with the $1-2$ 
mixing angle given by, 
\be
\label{s1}
\tan 2 \th_1 =  {2 B'_{12} \over B'_{22} - B'_{11}} 
\ee

In the above,  we have made no assumption on the relative magnitudes of 
$m_0$ and $m_\la$. Only assumption is  almost equality of all the 
$\l$ couplings.  Most definitions are formally the same as in case of 
the $\l'$ couplings but the physical consequences are quite different.
This follows from the expressions of mixing angles $s_{2,3}$ in
eq.(\ref{s2s3}).
If  no hierarchy is assumed among the $\l$ couplings then due to antisymmetry
of $\l$ we get,

$$ {b_2\over b_3}\approx {b_1\over b_3}\approx {\cal O} ({h_2^{E~2}\over h_3^{E~2}}) $$

As a result, eq.(\ref{s2s3}) implies 
$c_3\sim {\cal O} ({m_\m\over m_\tau})$ and the mixing
matrix, eq.(\ref{mix}) can be written as

\bea \label{totalmix}
K &=&U~_{\l}U'_{\l}\nonumber \\
&=&\left( \ba{ccc}
c_1 c_2 &  s_1 c_2 & s_2 \\
-s_2 c_1  &- s_1 s_2  & c_2  \\
s_1  &  c_1 & 0 \ea \right)  +{\cal O}({m_\m\over m_{\tau}}).
\eea

The equations for $\eta_1, \eta_2$ and $\tan 2 \th_1 $ take the following
approximate forms in this case: 

\bea \label{etasapp}
\eta_1 &\approx & s_1^2 c_2^2 B_{11} + 2 B_{12} c_2 s_1 
(c_1 -s_1 s_2 ) \nonumber \\
\eta_2 &\approx& c_1^2 c_2^2 B_{11} - 2 B_{12} c_1 s_2 (s_1 + c_1 s_2)
\eea

\be
\label{s1app}
\tan 2 \th_1 \approx  {-2 B_{12} \over B_{11} c_2 - 2 s_2 B_{12}}
\ee 

With the hierarchical masses, the effective mixing angles $\theta_A$ and 
$\theta_{CHOOZ}$ probed by the atmospheric data and the CHOOZ experiment 
respectively are given by

$$ \sin^2 2 \theta_A=4 K_{\mu 3}^2 (1-K_{\mu 3}^2)~~~~~~~
~~~~~~~~~\sin^2 2 \theta_{CHOOZ}\approx 4 K_{e3}^2 (1-K_{e3}^2) ~.$$

Eq.(\ref{totalmix}) implies that these two mixing angles 
are equal in conflict with the observation.

The above derivation has not assumed any specific mechanism for 
the supersymmetry breaking and thus the conclusions are valid in 
supergravity scenario as well as in the MMM case considered here. 
It is quite interesting that in spite of the presence of nine 
independent $\l$ parameters, one cannot explain solar and atmospheric 
neutrino anomaly as long as these parameters are of similar magnitudes.
Departure from equalities of $\l$ can lead to explanations of these
anomalies and we shall present a specific example in the next section.

\subsection{MMM and Neutrino Anomalies}

In this sub-section, we first determine the numerical values of the parameters
$m_\la$ and $m_0$ which enter the neutrino mass. The dependence on
trilinear couplings is factored out in defining these parameters. 
But their numerical values are quite different here compared to 
the $\l'$ case studied in the earlier section.
This follows since the soft sfermion masses which determine these 
parameters are quite different in these two cases ( see RG equations 
in the appendix ).
 
In models with gauge mediated SUSY breaking, the soft masses 
are proportional to the gauge quantum numbers they carry. Thus particles 
with strong interactions have much larger soft masses compared to the 
weakly interacting  particles, as is evident from the 
eqs.(\ref{gmass},\ref{smass}). The effect of the gauge couplings
also trickles down to the parameters $B_{\e_i}, m^2_{\n_i H_1}$ through
the corresponding soft masses present in their respective renormalization
group equations. In the presence of purely $\l'$ interactions, the strong
coupling determines the magnitudes of $B_{\e_i}, m^2_{\n_i H_1}$ at the 
weak scale and in turn the tree level mass as is evident from  
eq.(\ref{apm0mmm}). The loop contribution is still however determined by the
weak coupling, eq.(\ref{apmlmmm}). It is thus the interplay between the strong
coupling and the weak coupling which lead to a suppressed loop mass in the case
of purely $\l'$ couplings.  In the case of pure $\l$ couplings, the squark 
masses do not enter the definition of $m_0$ and $m_\la$ and both of these 
are determined by the weak coupling. However the dependence on the power 
of the weak coupling is different. Following the same method as described 
in the $\l'$ case (above eq.(\ref{apm0mmm})), an estimate of the parameter
 $m_0$ is given by,

\be
\label{aplm0}
m_0 = \left( \cos \b \over 8 \pi^2 \right)^2 {M_W^2 \over \Lambda} 
{1 \over \tilde{\alpha}_2 (X)} \left( \ln {X^2 \over M_Z^2} \right)^2
\ee 

The 1-loop contribution $m_\la$ can also be estimated in the similar 
manner as,

\be
\label{aplml}
m_\la = \left( v^2 \m \over \Lambda^2 \right) { \cos \b \sin \b \over
24~ \pi^2 \tilde{\alpha}_2^2(X) }
\ee 

The ratio ${m_\la \over m_0}$ is then given by,
\be
\label{ratlam}
{m_\la \over m_0} \approx \left( { 8 \pi^2 \over 3 } \right) 
\left({ v^2 ~\m ~\tan \b \over t^2~ \Lambda~ M_W^2 } \right) 
{1 \over \tilde{\alpha}_2 (X)}
\ee 

where $ t = \ln \left( { X^2 \over M_Z^2} \right)$. The above is typically
of ${\cal O}(10^2)$ for $\Lambda = 100$ TeV, $\m = 400$ GeV, $\tan \b =46$ 
which shows that the tree level mass is much suppressed compared to the 
1-loop mass.  Comparing the above equation with that of the corresponding 
one for the $\l'$ case, eq.(\ref{apratiolamp}), we see that the absence of 
strong-weak interplay in this case leads to a much larger ratio.

In a general mSUGRA inspired scenario, the tree level
mass is much larger compared to the 1-loop mass \cite{sneutrino} for large
range in MSSM parameters, irrespective of the nature of $R$ parity breaking.
In the present case, the relative importance of loop and the tree
level contributions is sensitive to the nature of R violation as
demonstrated above. This feature arises not as a consequence of running 
of soft masses but due to difference in the relevant RG equations in
case of $\l$ and $\l'$ couplings and the boundary conditions
themselves which strongly depend on the gauge couplings in these models.

We have determined the ratio ${m_\la \over m_0}$ by solving the relevant
RG equations numerically in the manner described in section (3). 
This ratio is plotted versus $\Lambda$ in Fig.2 for $\Lambda=(50-150)$ TeV. 
In this range,

\be
{m_\la \over m_0} = 25 -45. 
\ee 
as expected from eq.(\ref{ratlam}). 

The dominance of $m_\la$ has the following important implication.
The neutrino mass ratio following from eq.(\ref{levalues}) is given by

\be
\label{egratiolam}
{m_{\nu_2} \over m_{\nu_3}} = {m_\la \over m_0 + m_\la} {m_\mu\over m_\tau} {\eta_2 
\over \eta_3} \ee
In the previous case, the hierarchy in $m_{\la}$ and $m_0$ resulted 
in strong hierarchy between neutrino masses. As a result, one could 
only obtain vacuum or quasi-vacuum solution for the solar neutrino. 
Here due to $m_\la\gg m_0$, hierarchy in neutrino mass is much weaker,
and we have,
\be
\label{ratlam1}
{m_{\nu_2} \over m_{\nu_3}} \approx {m_\mu \over m_\tau} \sim 5 \times 10^{-2}
\ee
We can thus easily get the scale relevant for the LMA solution of the 
solar neutrino.  As already argued the mixing pattern is not appropriate 
if all the $\l$ couplings are similar.  This is no longer true if $\l$ obey 
some specific hierarchy as we discuss now.

\subsection{Illustrative Model }

We assume that the couplings  $\l_{123}, \l_{233}, \l_{322}$ dominate 
over the rest and neglect the latter. Moreover we assume that non-zero 
couplings satisfy the following hierarchy,

\be
\label{hier}
{\l_{233} \over  \l_{322}} \approx{\cal O} \left({ h_2^E \over  h_3^E} 
\right) \;\;;\;\; 
{\l_{123} \over \l_{322} }\leq{\cal O} \left({ h_2^E \over  h_3^E} \right)
\ee

We do not have strong theoretical reasons to assume the above hierarchy.
The following considerations should therefore be viewed as an example which
leads to the successful explanation of the neutrino anomalies.

The tree level mass matrix in the presence of these couplings is given 
by,
\be
{\cal M}_s^0 = m_0 \bmat{ccc}
0 &0&0 \\
0& \l_{233}^2 h^{E~2}_3& \l_{233} h_3^E \l_{322} h_2^E \\
0& \l_{233} h_3^E \l_{322} h_2^E & \l_{322}^2 h^{E~2}_2 \emat
\ee 
The 1-loop level mass matrix is given by, 
\be
{\cal M}_s^l = m_\la \bmat{ccc}
0& \l_{123} h_2^E \l_{232} h_3^E& 0 \\
\l_{123} h_2^E \l_{232} h_3^E & \l_{233}^2 h^{E~2}_3 & 
\l_{232} h_2^E \l_{323} h_3^E \\
0& \l_{232} h_2^E \l_{323} h_3^E & \l_{322}^2 h_2^{E~2}
\emat ~.
\ee
In view of the hierarchy in eq.(\ref{hier}), the total mass matrix has the 
following simple form
\be
{\cal M}_s \approx \bmat{ccc}
0 & x & 0 \\
x & A & A  \\
0 & A & A \emat
~,\ee 
where 
\bea
\label{xa}
A&\equiv& (m_0+m_\la) \l_{233}^2 h_3^{E~2} ~,\nonumber \\
x&\equiv&-m_\la \l_{322}\l_{123} h_2^Eh_3^E ~.\eea 
We shall assume $x\leq A$ which is consistent with the hierarchy in 
eq.(\ref{hier}).
One can diagonalise the above matrix:
$$ R_{12}(\theta_{12})~R_{13}(\theta_{13})~R_{23}(\pi/4)~{\cal M}_s
\left[R_{12}(\theta_{12})~R_{13}(\theta_{13})~R_{23}(\pi/4)\right]^T
\approx Diag.(m_{\nu_1},m_{\nu_2},m_{\nu_3}) ~.$$
Here, $R_{ij}$ denotes rotation in the $ij^{th}$ plane with angle 
$\theta_{ij}$. We have neglected a small contribution of 
${\cal O}({x\over 2\sqrt{2} A})$ 
to the $2-3$ mixing angle in the above derivation. The mixing angles are 
given by
\bea
\label{angles} 
\tan 2 \theta_{12}&\approx& {4 \sqrt{2} A\over x} ~, \nonumber \\
\tan 2 \theta_{13}&\approx& {x\over A\sqrt{2}}~.
\eea
The eigenvalues can be approximately written as
\bea
\label{masses} 
m_{\nu_1}&\approx& {x\over \sqrt{2}}-{x^2\over 8 A}~, \nonumber \\
m_{\nu_2}&\approx&- {x\over \sqrt{2}}-{x^2\over 8 A}~, \nonumber \\
m_{\nu_3}&\approx& 2 A+{x^2\over 2 A}~.
\eea

It is seen from the last two equations that all the mixing angles and 
the masses are predicted in terms of only two parameters namely,
$x$ and $A$. The atmospheric mixing angle is predicted to be around
$\pi/4$ and the other two mixing angles can be expressed in terms of the
solar and atmospheric scales. Using eqs.(\ref{angles},\ref{masses}) we find,

\bea
\label{scales}
\tan 2 \theta_{solar}&\approx& {4\sqrt{2}A\over x}\approx 
4\sqrt{2} \left({8\sqrt{2}\Delta_{solar}\over \Delta_A}\right)^{-1/3}
~,\nonumber \\
\tan 2 \theta_{CHOOZ}&\approx& {x\over A\sqrt{2}}\approx 
{1\over \sqrt{2}}\left({8\sqrt{2}\Delta_{solar}\over \Delta_A}\right)^{1/3}~. 
\eea
 
Choosing $\Delta_A\sim 3\cdot 10^{-3}\eV^2$ and $\Delta_{solar}\sim
(2-50) \cdot 10^{-6} \eV^2$ we get,
\bea
{x\over A}&\approx& (0.19-0.57)~, ~ \nonumber \\
\tan^2 \theta_{solar}&\approx& (0.93- 0.81)~,\nonumber \\
U_{e3} (CHOOZ)&\approx &(0.06-0.19)~. 
\eea
The predictions for the  mixing angles are in very good agreement with the observations
which prefer large mixing angle solution for the solar neutrino. 
The required value of
${x\over A}$ is also consistent with the assumed hierarchy in
eq.(\ref{hier}) among the
trilinear couplings. 

\section{Discussion}

We have discussed the structure of neutrino masses and mixing in the 
Minimal messenger model (MMM) of gauge mediated supersymmetric breaking
with purely trilinear R violating interactions. We considered two 
specific cases of purely $\l'$ interactions and purely $\l$ interactions
for simplicity.  The model contains very large number
of parameters even under this simplifying assumptions. Remarkably, 
it is possible to make meaningful statement on the 
neutrino spectrum in spite of the presence of many unknown parameters
if all these parameters are assumed similar in magnitude. This is a 
natural assumption in the absence of any specific symmetry to restrict 
the trilinear $R$ parity violation. It is not always easy to justify
this specific choice, e.g use of a $U(1)$ symmetry which uses Froggatt
-Nielsen [FN] mechanism to obtain quark and lepton masses tend to forbid
all the trilinear terms altogether \cite{u1paper}. 

In the case where only $\l'$ couplings are present, one naturally gets 
large mixing between the neutrino states. Further, 
the MMM  offers a very constrained structure giving rise to a large 
hierarchy between the masses $\sim O(10^{-2})$ for all the parameter space. 
The model  is suitable
for obtaining simultaneously solutions for atmospheric neutrino problem
and quasi-vacuum oscillations. 

Assumption of approximate equality of $\l$ couplings in case with only 
$\l$ couplings, leads to very constrained and phenomenologically 
inconsistent pattern for neutrino mixing. This conclusion
follows on general grounds and it is true even if SUSY breaking is induced
by supergravity interactions. 
It is quite interesting that one can arrive at this strong conclusions 
in spite of the presence of many unknown parameters by simply assuming 
them to be of similar magnitude. 

One can obtain consistent picture of neutrino anomalies if $\l$ couplings 
are assumed to be hierarchical. We provided an example which leads to two large 
and one small mixing and correct hierarchy between the solar and atmospheric 
neutrino scales.
 
One interesting result of this analysis is the interplay between the 
sneutrino {\it vev} induced contribution and the loop induced contribution to 
neutrino masses. In the context of supergravity induced SUSY breaking, it
has been shown that the large logarithmic factors induced due to RG scaling
enhance the sneutrino {\it vev} induced contribution compared to 
the loop contribution. We showed that this remains true even in the gauge
mediated models of SUSY breaking in case of the trilinear $\l'$ 
couplings. In the mSUGRA model, the dominance of tree level mass
follows simply from the large factor $t = ln\frac{M^2_{GUT}}{M^2_{Z}}$ 
in sneutrino {\it vev} generated by running of soft parameters. In the 
present case, the tree level dominance occurs essentially due to
boundary conditions. In case of $\l'$ couplings, $m_0$ is determined
by squark masses which depend upon $\a_3 (X)$. $m_0$ dominates over
loop contribution in this case. For $\l$ couplings, $m_0$ is
determined by sleptons rather than by squarks masses. Due to their
dependence on weak couplings, slepton masses are much smaller than
squark masses. As a consequence, $m_0$ is suppressed compared to $\l'$ 
case. This results in loop dominance if R is violated by $\l$ couplings.

The neutrino mass hierarchy strongly depends on the ratio $\frac{m_{\la}}{m_0}$.
In case of the tree level dominance (purely $\l'$ couplings) one obtains 
strong hierarchy and vacuum solution while the case with loop mass dominating 
corresponds to milder hierarchy and the LMA MSW solution.  This analysis
along with other similar analysis \cite{sneutrino,carlos,chun} therefore 
underlines the need of including both the contributions to neutrino masses
in a proper way.

\noindent
{\bf Acknowledgments:}
We warmly acknowledge discussions with U. Chattopadhyay which have been
very helpful. 

\begin{center}
{\bf Appendix I}
\end{center}

Here we present the Renormalization Group Equations (RGE) for the soft
parameters $B_{\e_i}$ and $m^2_{\n_i H_1}$ for the two cases considered
in this work: either purely $\l'$ couplings or purely $\l$ couplings are 
the sources of lepton number violation in the superpotential. These equations 
have been derived using the general formulae given in \cite{falck}. These 
equations can also be found in \cite{carlos}, whereas the equations for 
the standard soft parameters appearing in the RHS of the equations can be 
found in many papers such as \cite{mssmrge}. In writing the below 
eqs.(\ref{lprge})(eqs.(\ref{lrge})), we have neglected ${\cal O}(\l^{'~2})$
(${\cal O}(\l^2)$) corrections. The notation is as described in the text.

\noindent
{\it $\l'$ couplings in the superpotential:}

\bea
\label{lprge}
{d B_{\e_i}(t) \over dt}&=& B_{\e_i}(t) \left( -{3 \over 2} Y_t(t) - {1 \over 2}
Y_i^E(t) + {3 \over 2} \tilde{\a}_2(t) + {3 \over 10} \tilde{\a}_1(t) \right)
\nonumber \\
&-& {3 \over 16 \pi^2} \mu(t) \l'_{ijj}(t)
h^d_{jj}(t) \left( {1 \over 2} B_\m(t) + A^{\l'}_{ijj}(t) \right) \nonumber \\
{d m_{\n_i H_1}^2 (t)\over dt}&=& m_{\n_i H_1}^2(t) \left( - {1 \over 2} 
Y^E_i(t)- {3 \over 2} Y_b(t) - {1 \over 2} Y_\tau(t) \right) - 
{ 3 \over 32 \pi^2 } \l'_{ipp}(t) h^d_{pp}(t)
\left( m_{H_1}^2(t) \right. \nonumber \\
&+& \left. m_{L_i}^2(t) + 2~ m_{Q_p}^2(t) + 2~ A_{ipp}^{\l'}(t)
 A^D_{pp}(t) + 2~ m^2_{D_p}(t) \right)  
\eea

\noindent
{\it $\l$ couplings in the superpotential: }

\bea
\label{lrge}
{d B_{\e_i}(t) \over dt}&=& B_{\e_i}(t) \left( -{3 \over 2} Y_t (t) - 
{1 \over 2} Y^E_i(t) + {3 \over 2} \tilde{\a}_2(t) + {1 \over 2} 
\tilde{\a}_1(t) \right) \nonumber\\
&-& {1 \over 16 \pi^2}  \m(t) \l_{idd}(t) h^E_{dd}(t) \left( A^\l_{idd}(t) +
{1 \over 2} B_\m(t) \right) \nonumber \\
{ d m^2_{\n_i H_1}(t) \over dt}&=&m^2_{\n_i H_1}(t) \left( - {1 \over 2} 
Y^E_i(t) - {1 \over 2} Y_\tau(t) - {3 \over 2} Y_b(t) \right) - 
{1 \over 32 \pi^2} \l_{ijj}(t) h^E_{jj}(t) \left( m_{H_1}^2(t) \right. 
\nonumber \\
&+& \left. m_{L_i}^2(t) + 2~ m_{L_j}^2(t) + 2~ A^{\l}_{ijj}(t) A^E_{jj}(t)
+ 2~ m_{E_j}^2(t) \right) 
\eea

\newpage
\begin{figure}[h]
\centerline{\psfig{figure=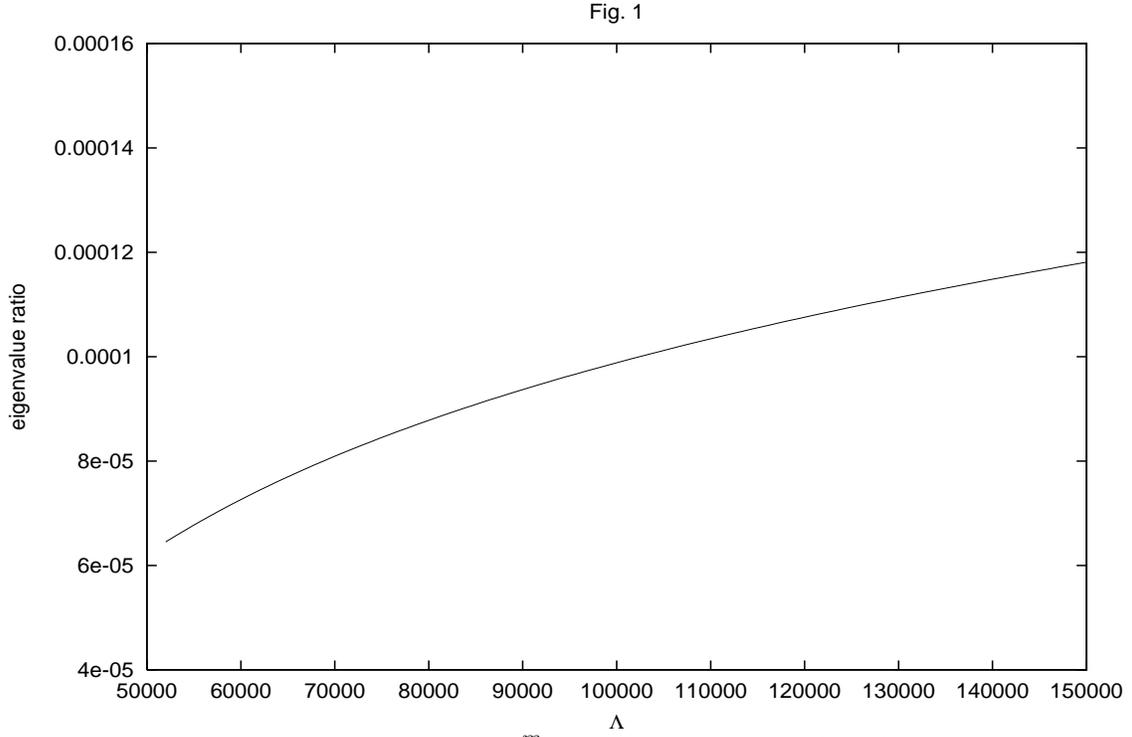,height=10cm,width=15cm,angle=-90}}
\caption{Neutrino mass eigen value ratio 
${m_{\n_2} \over m_{n_3}}$ plotted versus $\Lambda$ (GeV) assuming trilinear 
$\l'$ couplings of similar strength to be the only source of $R$ violation. 
$\Lambda$ is defined in the text. }
\end{figure}

\newpage
\begin{figure}[h]
\centerline{\psfig{figure=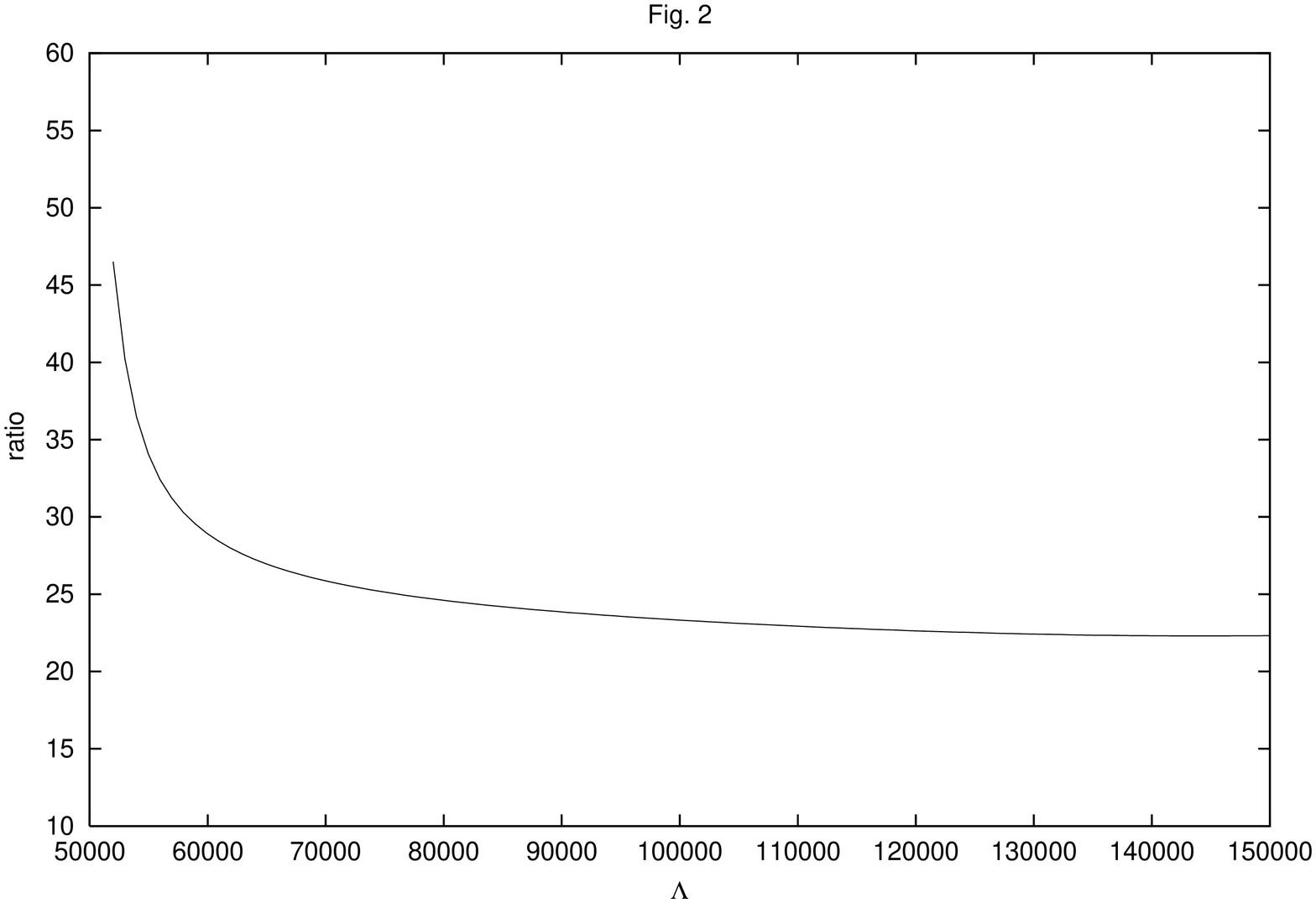,height=10cm,width=15cm,angle=-90}}
\caption{The neutrino mass ratio 
${m_{loop} \over m_0}$ plotted versus $\Lambda$ (GeV) assuming  trilinear $\l$
couplings of similar strength to be the only source of $R$ violation.}
\end{figure}


\begin{thebibliography}{99}
%
\bibitem{skatm} 
S.~Fukuda {\it et al.}  [Super-Kamiokande Collaboration],
Phys.\ Rev.\ Lett.\  {\bf 85} (2000) 3999
%
\bibitem{sksolar} 
S.~Fukuda {\it et al.}  [Super-Kamiokande Collaboration],
hep-ex/0103033.
%
\bibitem{choose} 
M.~Apollonio {\it et al.}  [CHOOZ Collaboration],
Phys.\ Rev.\ D {\bf 61} (2000) 012001
[hep-ex/9906011].
%
\bibitem{revs} For reviews of various models, please see,  
A. Yu. Smirnov, hep-ph/9901208; R. N. Mohapatra, hep-ph/9910365; 
A. S. Joshipura, Pramana, {\bf 54} (2000) 119.  
%
\bibitem{hall}
L. J. Hall and M. Suzuki, \np{B 231}{84}{419}. 
%
\bibitem{asjbabu} 
K. S. Babu and A. S. Joshipura, Talks at the 
Trieste Conf. on `Quarks and Leptons: Masses and Mixing', Oct.1996, Oct,1997.
%
\bibitem{asjskv}
A. S. Joshipura and S. K. Vempati, \pr{D60}{99}{095009}.
%
\bibitem{bilinear_univ}
T.Banks et. al, \pr {D 52}{95}{5319};
R. Hempfling, \np{B478}{96}{3}; 
H. Nilles and N. Polonsky \np{B484}{97}{33}; 
M. A. Diaz, J. C. Romao and J. W. F. Valle, \np{B524}{98}{23}.
E. J. Chun, S. K. Kang, C. W. Kim, \np{B544}{99}{89}; E. J. Chun and
J. S. Lee, \pr {D 60}{99}{075006}; 
E. J. Chun, {\it Phys. Lett.} {\bf B505}(2001) 155.
%
\bibitem{bilinear_nouniv}
A. Yu. Smirnov and F. Vissani, \np{B460}{96}{37};
D. Du and C. Liu, \mpl {A10}{95}{1837}; C. Liu, \mpl {A12}{97}{329};
S. Roy and B. Mukhopadhyaya, \pr{D55}{97}{7020}; 
B. Mukhopadhyaya, S. Roy and F. Vissani, \pl{B443}{98}{191}; 
G. Dvali and Y. Nir, JHEP {\bf 9810:014},1998 ;
M. Bisset, O. C. W. Kong, C. Macesanu, L. H. Orr, Phys. Rev. {\bf D 62}
(2000) 035001. 
%
\bibitem{romao}
M. Hirsch, M. A. Diaz, W. Porod, J. C. Romao and J. W. F. Valle, 
Phys. Rev. {\bf D 62} (2000) 113008. Also see, by the same authors, 
{\it Phys. Rev.}{\bf D61} (2000) 071703; hep-ph/0009066; 
hep-ph/0009127. 
%
\bibitem{kaplan}
D. E. Kaplan and A. E. Nelson, JHEP {\bf 033:0001}, 2000. 
%
\bibitem{rhneut} 
R. Kitano and K. Oda, {\it Phys. Rev} {\bf D 61} (2000) 113001. 
%
\bibitem{abada}
S. Rakshit, G. Bhattacharyya and A. Raychaudhuri, \pr {D59}{99}{091701}.
A. Abada and M. Losada, Nucl. Phys. {\bf B 585} (2000) 45; 
R. Adhikari and G. Omanovic, hep-ph/9802390; \pr {D 59}{99}{073003};
O. Haug, J. D. Vergados, A. Faessler and S. Kovalenko, {\it Nucl. Phys.}
{\bf B 565} (2000) 38; R. Adhikari, A. Sil and A. Raychaudhuri, hep-ph/0105119. 
%
\bibitem{drees} 
M. Drees, S. Pakavasa, X. Tata and T. ter Veldhuis,  \pr{D 57}{98}{R5335}.
%
\bibitem{sneutrino} 
A. S. Joshipura and S. K. Vempati, \pr{D60}{99}{111303}. 
%
\bibitem{carlos} 
B. de Carlos and P. L. White, \pr{D54}{96}{3427};
E. Nardi,\pr{D55}{97}{5772}.  
%
\bibitem{gamberini}
G. Gamberini, G. Ridolfi and F. Zwirner, \np{B331}{90}{331}. 
%
\bibitem{chun1loop}
E. J. Chun, S. K. Kang, {\it Phys. Rev.}~{\bf D61}(2000) 075012; 
E. J. Chun, hep-ph/0001236.
%
\bibitem{borzu} 
F. Borzumati, hep-ph/9702307.
%
\bibitem{asjmarek} 
A. S. Joshipura and M. Nowakowski, \pr{D51}{95}{2421}.
%
\bibitem{dine} 
M. Dine, A. E. Nelson and Y. Shirman, \pr {D 51}{95}{1362}; 
M. Dine, A. E. Nelson, Y. Nir and Y. Shirman, \pr{D 53}{96}{2658}.
For original references and  reviews please see, 
G. F. Giudice and R. Rattazzi, \prep{322}{99}{419};
S. L. Dubovsky, D. S. Gorbunov and S. V. Troitsky, 
{\it Phys. Usp.} {\bf 42} (1999) 623, [hep-ph/9905466]. 
%
\bibitem{chun}
K. Choi, K. Hwang and E. J. Chun, \pr {D 60}{99}{031301} and 
Ref. \cite{chun1loop}.
%
\bibitem{mmm}
M. Dine, Y. Nir and Y. Shirman, \pr {D 55}{97}{1501};
K. S. Babu, C. Kolda and  F. Wilczek, \prl{77}{96}{3070};
S. Dimopoulos, S. Thomas and J. D. Wells, \np{B488}{97}{39};
J. A. Bagger, K. Matchev, D. M. Pierce and R. Zhang, \pr{D55}{97}{3188}. 
%
\bibitem{martin}
S. P. Martin , \pr{D 55}{97}{3177}
%
\bibitem{rattazzi}
U. Sarid and R. Rattazzi, \np {B 501}{97}{297}.

\bibitem{hera1}
A. S. Joshipura, V. Ravindran and S. K. Vempati, \pl{B451}{99}{98}.
%
\bibitem{aleph}
R. Barate {\it et. al}, [ALEPH Collaboration], {\it Eur. Phys.} {\bf J ~C 16}
(2000) 71. 
%
\bibitem{gonzalez}
M.~C.~Gonzalez-Garcia, M. Maltoni, C. Penya-Garay and J. W. F. Valle,
Phys. Rev. {\bf D 63} (2001)(033005); J. Bahcall, P. I. Krastev and
A. Yu. Smirnov, {\it JHEP}~{\bf 0105:015}(2001). 

%
\bibitem{oyama} 
Y. Oyama {\it et. al}, [Super-Kamiokande Collaboration], hep-ex/0104015.
%
\bibitem{sno}
Two generation analyses including the recent results from SNO (nucl-ex/0106015) have
been presented in, G. L. Fogli {\it et. al}, hep-ph/0106247; J. N. Bahcall {\it et. al},
hep-ph/0106258; A. Bandyopadhyay {\it et. al}, hep-ph/0106264. Also see, V. Barger 
{\it et. al}, hep-ph/0106207.
%
\bibitem{falck} 
N. K. Falck, \zp{C30}{86}{247}. 
%
\bibitem{mssmrge} For a list of MSSM RGE, please see, 
L. E. Ibanez and C. Lopez, \np {B233}{84}{511}; M. Carena {\it et. al}
\np {B 491}{97}{103}. 
%
\bibitem{u1paper}
A. S. Joshipura, R. D. Vaidya and S. K. Vempati, {\it Phys. Rev}~ 
{\bf D 62} (2000) 093020. 
%
\end{thebibliography}
\end{document}